\documentclass[aip,twocolumn, showpacs,superscriptaddress,graphicx]{revtex4}
\usepackage{amssymb}
\usepackage{graphicx}
\usepackage{amsmath}
\usepackage{amsfonts}
\usepackage{epsfig}
\usepackage{epstopdf}

\draft

\begin{document}

\title{Anomalous surface states at interfaces in p-wave superconductors}
\author{S.~V.~Bakurskiy}
\affiliation{Faculty of Physics, Lomonosov Moscow State University, 119992 Leninskie
Gory, Moscow, Russia}
\affiliation{Moscow Institute of Physics and Technology, Dolgoprudny, Moscow 141700,
Russia}
\affiliation{Faculty of Science and Technology and MESA+ Institute for Nanotechnology,
University of Twente, 7500 AE Enschede, The Netherlands}
\author{A.~A.~Golubov}
\affiliation{Faculty of Science and Technology and MESA+ Institute for Nanotechnology,
University of Twente, 7500 AE Enschede, The Netherlands}
\affiliation{Moscow Institute of Physics and Technology, Dolgoprudny, Moscow 141700,
Russia}
\author{M.~Yu.~Kupriyanov}
\affiliation{Lomonosov Moscow State University Skobeltsyn Institute of Nuclear Physics
(MSU SINP), Leninskie gory, GSP-1, Moscow 119991, Russian Federation}
\affiliation{Moscow Institute of Physics and Technology, Dolgoprudny, Moscow 141700,
Russia}
\author{K.~Yada}
\affiliation{Department of Applied Physics, Nagoya University, Nagoya, 464-8603, Japan}
\author{Y.~Tanaka}
\affiliation{Department of Applied Physics, Nagoya University, Nagoya, 464-8603, Japan}
\affiliation{Moscow Institute of Physics and Technology, Dolgoprudny, Moscow 141700,
Russia}
\date{\today }
\date{\today }

\begin{abstract}
We present the results of theoretical study of surface state properties in a
two-dimensional model for triplet $p$-wave superconductors. We derive
boundary conditions for Eilenberger equations at rough interfaces and
develop the approach for self-consistent solution for the spatial dependence
of $p_{x}$ and $p_{x}+i~p_{y}$ -wave pair potentials. In the $p_{x}$ case we
demonstrate the robustness of the zero-energy peak in the density of states
(DoS) with respect to surface roughness, in contrast to the suppression of
such a peak in the case of $d_{xy}$ symmetry. This effect is due to
stability of odd-frequency pairing state at the surface with respect to
disorder. In the case of the chiral $p_{x}+i~p_{y}$ state we demonstrate the
appearance of a complex multi-peak subgap structure in the spectrum with
increasing surface roughness.
\end{abstract}

\maketitle

\section{Introduction\label{Intro}}

Investigation of spin-triplet superconductivity is currently an exciting
topic of research by the superconducting community. There are several
experimental results in Sr$_{2}$RuO$_{4}$ \cite%
{Maeno,Ishida,Luke,Mackenzie,Nelson,Asano} and in heavy fermion compounds 
\cite{Tou,Muller,Qian,Abrikosov,Fukuyama,Lebed,Saxena,Peiderer, Aoki} that
are consistent with spin-triplet superconducting pairing. The promising
paring symmetries are believed to be $p$-wave and $f$-wave in Sr$_{2}$RuO$%
_{4}$ \cite{Maeno,Ishida,Luke,Mackenzie,Kikugawa} and UPt$_{3}$ \cite{Graf,
Machida, Lussier}, respectively. Furthermore, to design $p$-wave
superconductivity based on a proximity coupled system with a conventional $s$%
-wave superconductor and the semiconductor surface state of a topological
insulator \cite{Alicea1,
Sau1,Sau2,Yamakage,Fu,akhmerov09,tanaka09,Law,Lutchyn,Oreg} has become a hot
topic from the viewpoint of topological superconductivity \cite%
{Ryu,Aliceareview,Qi,Nagaosa}. The essential ingredients in these new
systems are momentum-spin locking due to spin-orbit coupling and time
reversal symmetry breaking by an external field.

In the above systems, it is known that surface Andreev bound state (SABS) 
\cite{Buchholtz,Hara,Hu,Kashiwaya, Yada} is generated inside the energy gap
and stems from the topological properties of the bulk Hamiltonian \cite%
{Sato2011}. The SABS has become a prominent concept since the debate over
the pairing symmetry of high temperature superconductors (HTSs) \cite%
{Hu,Tanaka1}. In HTSs, if the angle between the direction normal to the
surface and the lobe direction of the $d$-wave pair potential deviates from
zero, the injected quasiparticle and the reflected one can feel opposite
signs of the pair potential depending on the injection angle \cite{Tanaka1}.
The extreme case is that the above angle becomes $\pm \pi /4$, where an
injected quasiparticle always feels the sign change independent of the
injection direction. This sign change of the pair potential produces SABS at
zero energy and induces the zero bias conductance peak in tunneling
spectroscopy \cite{Tanaka1,Kashiwaya,Lofwander}. The SABS has a flat
dispersion along $k_{y}$, where $k_{y}$ is the momentum parallel to the
surface. Actually, there are many experimental reports supporting ZBCP
stemming from SABS \cite%
{Kashiwayaexp,Covington,Alff,Wei,Iguchi,Biswas,Chesca}.

When the zero energy SABS is located at the surface or interface,
suppression of the pair potential in the main pairing channel occurs \cite%
{Nagato,Buchholtz95,Barash,Tanuma98}. Furthermore, if the time reversal
symmetry breaking is induced by the surface subdominant pair potential, ZBCP
can split \cite{Matsumoto95,Fogelstrom,Covington,Tanuma01}. Thus,
experimental study of the properties of ZBCP can serve as a guide to
determine the symmetry of the pair potential and the possible presence of a
subdominant one near the surface.

At the actual surface or interface, the diffusive scattering by the
roughness due to atomic scale irregularity inevitably exists. It is known
that surface roughness influences the electronic states of unconventional
superconductors such as those of $d$-wave or $p$-wave type \cite%
{Buchholtz86,WZhang,NagatoLTP}. Studies of conductivity at the interfaces in 
$d$-wave superconductors have shown that their properties are strongly
influenced by the degree of diffusive scattering of quasiparticles at the
interface \cite{Barash, Fogelstrom,Golubov1, Golubov1b, Golubov2}. The
higher the intensity of the diffusive scattering, the less pronounced the
conductance peak at low voltages and the more pronounced the influence of
subdominant components of order parameter on its shape.

Besides the above mentioned works, the theory of a proximity effect in
diffusive normal metal / $d$-wave superconductor junctions has been
developed \cite{Nazarov}. It has been clarified that SABS can not penetrate
into diffusive normal metal (DN) and the resulting ZBCP is broadened. These
properties can be naturally explained using the concept of odd-frequency
pairing \cite{Berezinskii}. The odd-frequency pairing states such as
spin-singlet $p$-wave or spin-triplet $s$-wave can be generated by the
translational symmetry breaking from the bulk conventional even-frequency
pairing state, $e.g.$, spin-singlet $s$($d$)-wave or spin-triplet $p$-wave 
\cite{Tanaka6,Eschrig07}. It is revealed that SABS in $d$-wave
superconductor might be interpreted as an odd-frequency spin-singlet $p$%
-wave pairing \cite{Tanaka6,Tanaka07}. However, $p$-wave pairing is fragile
against diffusive scattering, so it can not penetrate into DN metal. This
property is consistent with the fact that surface roughness has strong
effect on ZBCP and SABS in $d$-wave superconductor.

On the other hand, a recent study of SBAS in $p$-wave superconductors has
been stimulated by investigation of pairing symmetry in Sr$_{2}$RuO$_{4}$.
The existing theory of the proximity effect in spin-triplet $p$-wave
superconductors predicts that SABS produced by $p_{x}$-wave pairing can
penetrate into DN metal attached to a spin-triplet $p_{x}$-wave
superconductor \cite{Tanaka4}. This proximity effect induces many exotic
phenomena including a zero enegy peak in the local density of state (LDoS)
and negative local superfluid density \cite%
{Tanaka5,Tanaka2006,Asano2007,Fominov,Asano11,Higashitani13, Keles}. Since
the SBAS in a spin-triplet $p$-wave superconductor corresponds to
odd-frequency spin-triplet $s$-wave pairing, it is robust against impurity
scattering \cite{Tanaka07}. 

In actual Sr$_{2}$RuO$_{4}$, the promising symmetry is chiral $p$-wave
pairing, $i.e.$, $p_{x}+ip_{y}$ and one can expect more complex state as
compared to $p_{x}$-wave or $d_{xy}$-wave cases. The resulting SABS has a
linear dispersion as a function of $k_{y}$ \cite{Matsumoto99,Furusaki} which
is different from SABS in spin-singlet $d$-wave or spin-triplet $p_{x}$-wave
superconductor. For a ballistic junction without any roughness, it has been
shown that the resulting conductance exhibits a wide variety of line shapes
including broad ZBCP or dip like structure around zero voltage \cite%
{Yamashiro,Honerkamp,Sengupta, Laube}. Although it is not easy to obtain
reliable tunneling spectroscopy data in the $ab$-plane junction
experimentally, recent fabrication of well oriented junctions enabled
detection of the SABS \cite{Kashiwaya11}. However, the effect of diffusive
scattering has not been clarified yet. For a detailed comparison with
experiment and predicted surface state, the research in this direction is
needed. Since there are several relevant works in the surface state of
superfluid $^{3}$He \cite{Higashitani,Higashitani2}, it is currently a
challenging issue to study surface roughness effect on the surface density
of states (SDoS) and pairing symmetry of chiral $p$-wave superconductors.

Despite the fact that previous studies revealed important aspects of these
phenomena \cite{Zhang, Matsumoto2, Nagato3}, there is still a need for
systematic study and quantitative predictions. The purpose of this study is
to evaluate the influence of the degree of diffusive electron scattering at
interfaces in $p$-wave superconductors on the DoS.



The structure of this paper is the following: in Section \ref{Sec1} we
formulate the problem and derive effective boundary conditions for diffusive
surfaces in $p$-wave superconductors. In the following sections we discuss
microscopic properties of pairing in such systems for the cases of both $%
p_{x}$ and chiral $p_{x}+ip_{y}$ symmetry. In the Sec.\ref{Sec2} we focus on
the spatial dependence of pair potential $\Delta $; Sec.\ref{Sec3} is
devoted to pair amplitudes $f$ and finally in Sec.\ref{Sec4} we consider DoS
for various surface properties.

\section{Model\label{Sec1}}

The description of the suppression of superconductivity in the main pairing
channel and of the generation of subdominant order parameters can be done
within the framework of the quasiclassical Eilenberger equations \cite%
{Eilenberger} within a two-dimensional model. To solve the problem, we will
assume that the conditions of the clean limit are valid in the bulk
superconductor region (scattering time $\tau \rightarrow \infty $ ) and the
equations have the form

\begin{equation}
2\omega f(x,\theta )+v\cos (\theta )\frac{d}{dx}f(x,\theta )=2\Delta
g(x,\theta ),  \label{El0}
\end{equation}%
\begin{equation}
2\omega f^{+}(x,\theta )-v\cos (\theta )\frac{d}{dx}f^{+}(x,\theta )=2\Delta
^{\ast }g(x,\theta ),  \label{El0a}
\end{equation}%
\begin{equation}
2v\cos (\theta )\frac{d}{dx}g_{\omega }(x,\theta )=2\left( \Delta ^{\ast
}f_{\omega }-\Delta f_{\omega }^{+}\right) .  \label{El0b}
\end{equation}%
Here $g(x,\theta )$, $f(x,\theta )$ and $\ f^{+}(x,\theta )$ are normal and
anomalous Eilenberger functions, $\Delta (x)$ is pair potential, $\theta $
is angle between the vector normal to the interface and the direction of the
electron Fermi velocity $v;$ $\omega =\pi T(2n+1)$ are Matsubara frequencies
and $T$ is temperature, $x$ is coordinate along the axis normal to the
boundary. The form of self-consistency equation is sensitive to the chosen
symmetry of pair potential. In the case of $p_{x}$-wave pairing potential $%
\Delta =\Delta _{x}\cos (\theta )$ leads to equation 
\begin{equation}
\Delta _{x}\ln \frac{T}{T_{c}}+2\pi T\sum_{\omega }\frac{\Delta _{x}}{\omega 
}-\left\langle 2\cos (\theta ^{\prime })~f(x,\theta ^{\prime })\right\rangle
=0.  \label{Sc0a}
\end{equation}%
The other type of chiral $p_{x}+i~p_{y}$ symmetry relates to $\Delta =\Delta
_{x}\cos (\theta )+i~\Delta _{y}\sin (\theta )$ with similar self-consistent
equation \cite{Bruder}

\begin{eqnarray}
\Delta _{x}\ln \frac{T}{T_{c}}+2\pi T\sum_{\omega }\frac{\Delta _{x}}{\omega 
}-\left\langle 2\cos (\theta ^{\prime }) Re f(x,\theta ^{\prime
})\right\rangle &=&0,  \label{Sc0b} \\
\Delta _{y}\ln \frac{T}{T_{c}}+2\pi T\sum_{\omega }\frac{\Delta _{y}}{\omega 
}-\left\langle 2\sin (\theta ^{\prime }) Im f(x,\theta ^{\prime
})\right\rangle &=&0.  \label{Sc0c}
\end{eqnarray}

\begin{figure}[tbh]
\begin{center}
\includegraphics[width=8.5cm]{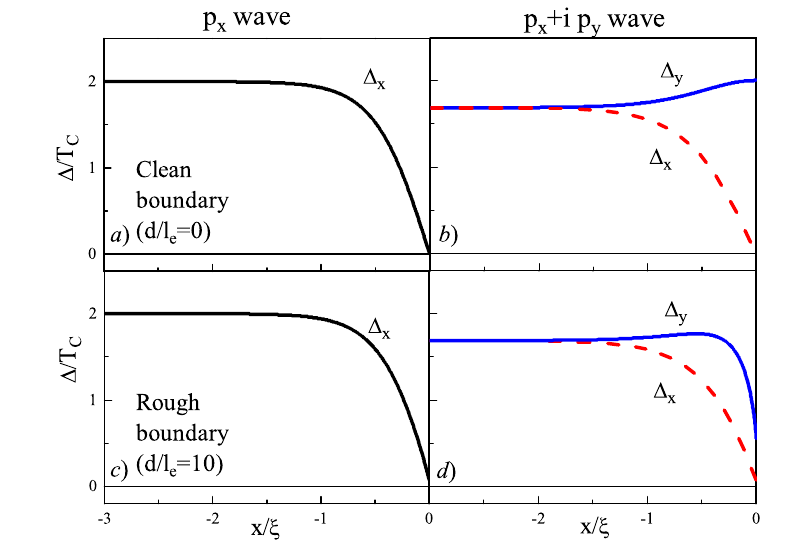}
\end{center}
\caption{Pair potential $\Delta $ as a function od coordinate $x$ in the
vicinity of surface for a) $p_{x}$-wave superconductor with clean surface,
b) $p_{x}+ip_{y}$-wave superconductor with clean surface, a) $p_{x}$-wave
superconductor with rough surface, a) $p_{x}+ip_{y}$-wave superconductor
with rough surface. On the panels b) and d) solid and dashed lines
correspond to components $\Delta _{y}$ and $\Delta _{x}$ respectively. Tha
parameter $d/l_{e}$ determines the degree of the surface roughness. }
\label{Del_x}
\end{figure}

Here $\left\langle ...\right\rangle =(1/2\pi )\int_{0}^{2\pi }(...)d\theta $
and $T_{c}$ is the critical temperature. Note that the considered case of
chiral p-wave supercondutor is equivalent to a thin film of superfluid 3He
A-phase. The polar phase of 3He has been recently identified in aerogel \cite%
{3He1,3He2}.

Diffusive properties of the interface will be described in the Ovchinnikov
model \cite{Ovchinnikov} , i.e. it is simulated by a thin diffusive layer of
thickness, $d\ll \xi _{eff}=\min \left\{ \sqrt{\xi _{0}l_{e}},\ \xi
_{0}\right\} ,$ $\xi _{0}=v/2\pi T_{c},$ with strong electron scattering
inside. Here $l_{e}$ is electron mean free path and $\tau =l_{e}/v$. Inside
this layer, located in the area $0\leq x\leq d$, we can neglect terms in the
Eilenberger equations \cite{Eilenberger} that are proportional to $\omega $
and $\Delta $ 
\begin{equation}
v\cos (\theta )\frac{d}{dx}f(x,\theta )=\frac{1}{\tau }\left( g\left\langle
f\right\rangle -f\left\langle g\right\rangle \right) ,  \label{El2a}
\end{equation}%
\begin{equation}
v\cos (\theta )\frac{d}{dx}f^{+}(x,\theta )=\frac{1}{\tau }\left(
g\left\langle f^{+}\right\rangle -f^{+}\left\langle g\right\rangle \right) ,
\label{El2b}
\end{equation}%
\begin{equation}
2v\cos (\theta )\frac{d}{dx}g(x,\theta )=\frac{1}{\tau }\left( f\left\langle
f^{+}\right\rangle -f^{+}\left\langle f\right\rangle \right)  \label{El2c}
\end{equation}%
and assume that $\left\langle f\right\rangle $, $\left\langle
f^{+}\right\rangle $ and $\left\langle g\right\rangle $ are spatially
independent quantities, which should be determined selfconsistently during
the process of finding solutions of the system (\ref{El0})-(\ref{El2c}). For
the development of numerical algorithms for solving the Eilenberger
equations it is convenient to rewrite them using the Ricatti parametrization 
\cite{Schopohl,Tanaka6}. 
\begin{equation}
f_{\pm }=\frac{2a_{\pm }}{1+a_{\pm }b_{\pm }},\ f_{\pm }^{+}=\frac{2b_{\pm }%
}{1+a_{\pm }b_{\pm }},\ g_{\pm }=\frac{1-a_{\pm }b_{\pm }}{1+a_{\pm }b_{\pm }%
},  \label{Ric1}
\end{equation}%
that are defined in the angle $-\pi /2\leq \theta \leq \pi /2.$ Their
substitution into (\ref{El0})-(\ref{El2c}) leads to the general relations in
the form

\begin{eqnarray}
v\cos (\theta )\frac{d}{dx}a_{\pm } &=&\Delta \left[ 1-a_{\pm }^{2}\right]
\mp 2\omega a_{\pm }  \label{El_ric_cl_A} \\
v\cos (\theta )\frac{d}{dx}b_{\pm } &=&-\Delta \left[ 1-b_{\pm }^{2}\right]
\pm 2\omega b_{\pm }  \label{El_ric_cl_B}
\end{eqnarray}%
in the clean superconducting region and 
\begin{eqnarray}
v\cos (\theta )\frac{d}{dx}a_{\pm } &=&\pm \frac{1}{2\tau }\left[
\left\langle f\right\rangle -a_{\pm }^{2}\left\langle f^{+}\right\rangle
-2a_{\pm }\left\langle g\right\rangle \right]  \label{El_ric_dr_A} \\
v\cos (\theta )\frac{d}{dx}b_{\pm } &=&\mp \frac{1}{2\tau }\left[
\left\langle f^{+}\right\rangle -b_{\pm }^{2}\left\langle f\right\rangle
-2b_{\pm }\left\langle g\right\rangle \right]  \label{El_ric_dr_B}
\end{eqnarray}%
in the diffusive layer. The subscript $\pm $ indicates the direction of
motion along the trajectory towards the boundary $(+)$ or away from it $(-)$%
. For $x\rightarrow -\infty $ we have 
\begin{equation}
a_{\pm }=\pm \frac{\Delta }{\omega +\sqrt{\omega ^{2}+|\Delta |^{2}}},
\label{As0}
\end{equation}%
\begin{equation}
b_{\pm }=\pm \frac{\Delta ^{\ast }}{\omega +\sqrt{\omega ^{2}+|\Delta |^{2}}}%
,  \label{As1}
\end{equation}%
where $\Delta $ is the bulk value of pair potential.

Finally, the problem must be supplemented by boundary conditions at the free
surface of the diffusion layer $(x=d)$%
\begin{equation}
b_{+}(d,-\theta )=b_{-}(d,\theta ),  \label{BC1a}
\end{equation}%
\begin{equation}
a_{-}(d,-\theta )=a_{+}(d,\theta ).  \label{BC1b}
\end{equation}%
The boundary conditions (\ref{BC1a}), (\ref{BC1b}) differ significantly from
those used previously \cite{Golubov2} 
\begin{equation}
b_{+}(d,-\theta )=a_{+}(d,\theta ),  \label{Eq92d}
\end{equation}%
in the analysis of the influence of diffuse scattering on the
superconducting correlations in $d$-wave superconductors. Indeed, in the $d$%
-wave case the following relations 
\begin{equation}
b_{\pm }(x,\theta )=a_{\mp }(x,\theta ),  \label{sym}
\end{equation}%
hold, and then the conditions (\ref{BC1a}), (\ref{BC1b}) are reduced to the
relation (\ref{Eq92d}). As a result, further analysis in the $d$-wave case
was based not on four, but only on two Eilenberger functions. It should be
also pointed out that when writing conditions (\ref{BC1a}-\ref{BC1b}) we
essentially used not only the fact that the particle reflected from the free
surface must diffuse into the node with a opposite value of the order
parameter, but also the fact that its velocity should be directed into the
interior of superconductor. That is why in the right side of (\ref{BC1a})
there is a function $b_{-}(d,\theta )$, and there is not $a_{+}(d,\theta )$,
or some combination of them.

\begin{figure}[tbh]
\begin{center}
\includegraphics[width=8.5cm]{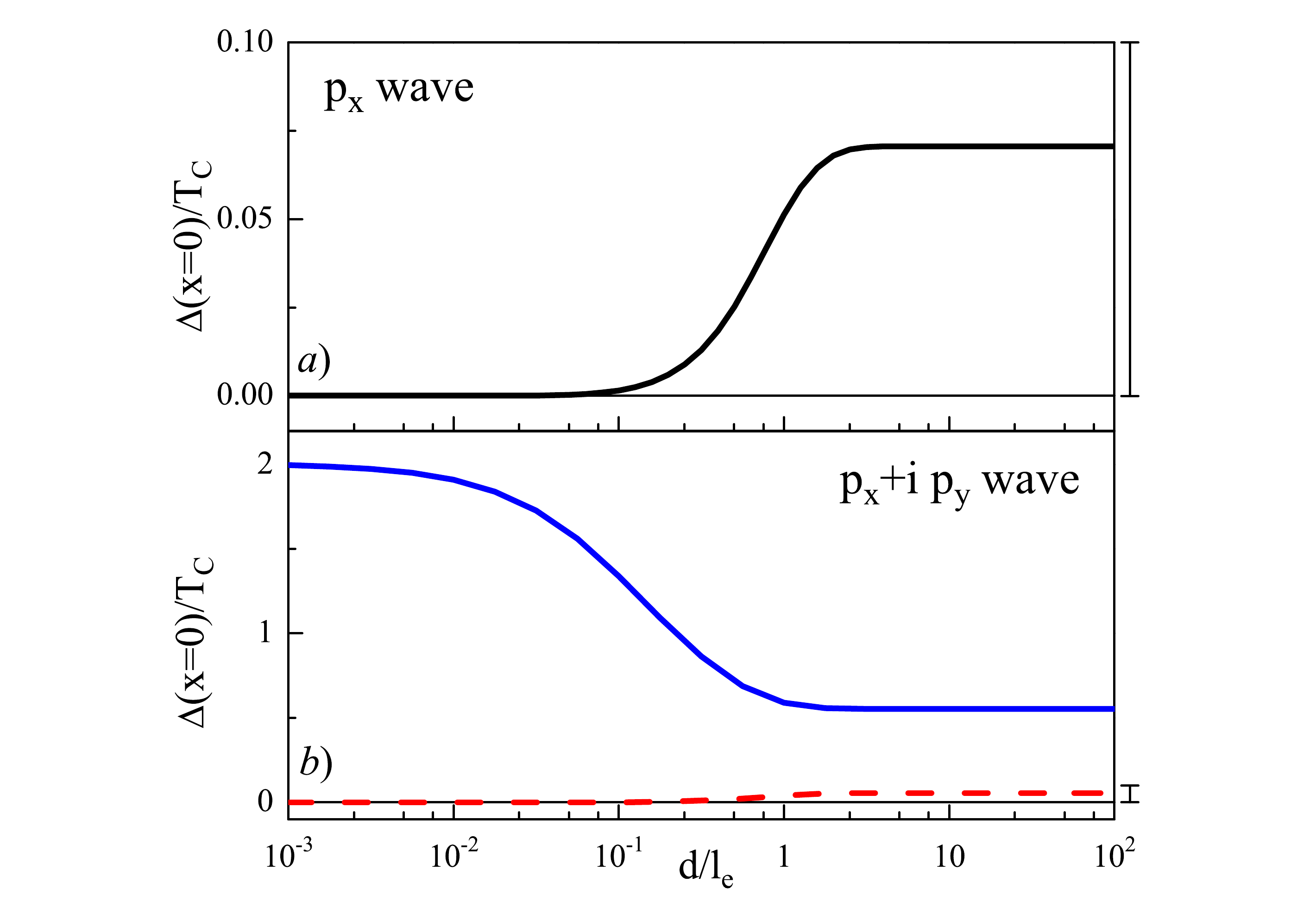}
\end{center}
\caption{Pair potential $\Delta $ on a surface of diffusive layer versus its
roughness $d/l_{e}$ a) for $p_{x}$-wave superconductor and b) $p_{x}+ip_{y}$%
-wave superconductor. Solid and dashed lines on the panel b) correspond to
components $\Delta _{y}$ and $\Delta _{x}$ respectively. }
\label{Del_dl}
\end{figure}

The boundary value problem (\ref{El_ric_cl_A})-(\ref{BC1b}) has been solved
analytically (see Appendix) resulting in an effective boundary condition at
the interface between the clean $p$-wave area and the diffusive layer at the 
$x=0$. It is expressed as the relation between the functions of a coming
into the diffusive layer $a_{+}(0,-\theta )$, $b_{-}(0,-\theta )$ and
leaving out from it $a_{-}(0,\theta )$, $b_{+}(0,\theta )$.%
\begin{equation}
a_{-}(0,\theta )=\frac{a_{+}(0,-\theta )-\left( Ga_{+}(0,-\theta )-F\right)
\tanh \left\{ kd\right\} }{\left( F^{+}a_{+}(0,-\theta )+G\right) \tanh
\left\{ kd\right\} +1},  \label{KG_A}
\end{equation}%
\begin{equation}
b_{+}(0,\theta )=\frac{b_{-}(0,-\theta )-\left( Gb_{-}(0,-\theta
)-F^{+}\right) \tanh \left\{ kd\right\} }{\left( Fb_{-}(0,-\theta )+G\right)
\tanh \left\{ kd\right\} +1}.  \label{KG_B}
\end{equation}%
Here $k$ is an effective wave vector in dirty layer 
\begin{equation}
k=\frac{\sqrt{\left\langle g_{+}+g_{-}\right\rangle ^{2}+\left\langle
f_{+}^{+}+f_{-}^{+}\right\rangle \left\langle f_{+}+f_{-}\right\rangle }}{%
\ell \cos (\theta )},
\end{equation}%
and $F$, $F^{+}$and $G$ are parametrized averages of Green functions%
\begin{eqnarray}
F &=&\frac{\left\langle f_{+}+f_{-}\right\rangle }{\sqrt{\left\langle
g_{+}+g_{-}\right\rangle ^{2}+\left\langle f_{+}^{+}+f_{-}^{+}\right\rangle
\left\langle f_{+}+f_{-}\right\rangle }},  \label{F_av1} \\
F^{+} &=&\frac{\left\langle f_{+}^{+}+f_{-}^{+}\right\rangle }{\sqrt{%
\left\langle g_{+}+g_{-}\right\rangle ^{2}+\left\langle
f_{+}^{+}+f_{-}^{+}\right\rangle \left\langle f_{+}+f_{-}\right\rangle }},
\label{F_av2} \\
G &=&\frac{\left\langle g_{+}+g_{-}\right\rangle }{\sqrt{\left\langle
g_{+}+g_{-}\right\rangle ^{2}+\left\langle f_{+}^{+}+f_{-}^{+}\right\rangle
\left\langle f_{+}+f_{-}\right\rangle }}.  \label{F_av3}
\end{eqnarray}%
Here the averaging operation is performed over the range of angles, $-\pi
/2\leq \theta \leq \pi /2,$ that is $\left\langle ...\right\rangle =(1/2\pi
)\int_{-\pi /2}^{\pi /2}(...)d\theta .$

The above boundary conditions are the main analytical result of this paper
and they provide the framework for a quantitative selfconsistent study of
surface effects in $p$-wave superconductors. The results of this study are
presented below.

\begin{figure}[tbh]
\begin{center}
\includegraphics[width=8.5cm]{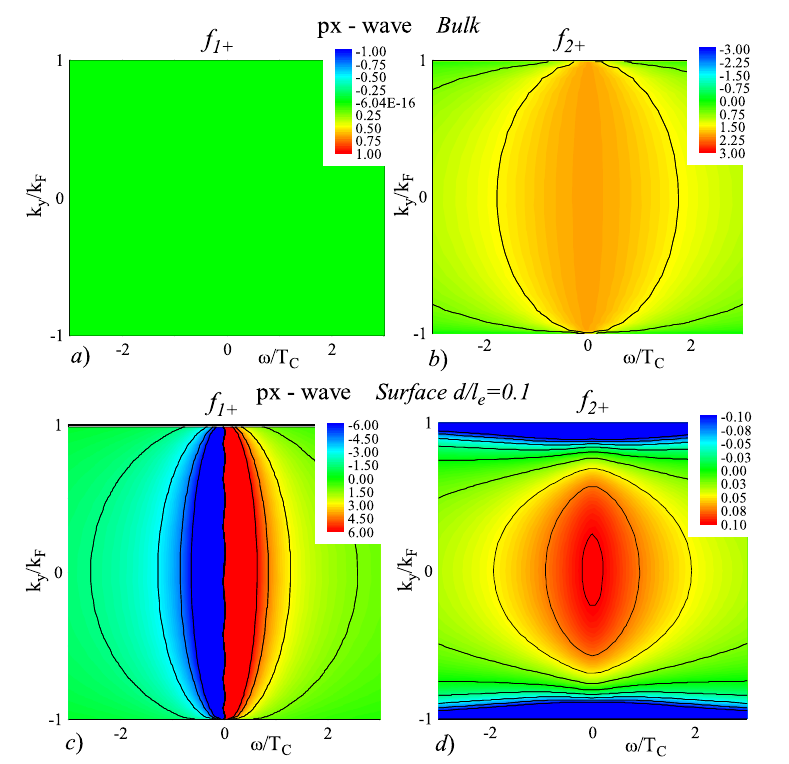}
\end{center}
\caption{(Color Online) Odd and even angle resolved pairing amplitudes $f_{1}
$ and $f_{2}$ as functions of $k_{y}$ and $\protect\omega $ a), b) in the
bulk $p_{x}$-wave superconductor and c),d) at the surface ($x=0$) with
roughness $d/l_{e}=0.1$ }
\label{f_px}
\end{figure}

\section{\protect\bigskip Pair potential, $\Delta $\label{Sec2}}

To study properties of the systems we have developed the method of numerical
solution of the boundary-value problem (\ref{El_ric_cl_A})-(\ref{BC1b}).
According to this method, outside of the diffusive layer (in the region
where $(\tau \rightarrow \infty )$), the equations (\ref{El_ric_cl_A}), (\ref%
{El_ric_cl_B}) for $a_{+}(x,\theta )$ and $b_{-}(x,-\theta )$ are
numerically integrated starting from conditions (\ref{As0}) at infinity $%
(x=-\infty )$ and moving along the trajectory towards the boundary $(x=0)$.
As a result, functions $a_{+}(0,\theta )$ and $b_{-}(0,-\theta )$ in
equations (\ref{KG_A}), (\ref{KG_B}) are calculated. Then, starting values $%
a_{-}(0,\theta )$ and $b_{+}(0,\theta )$ are determined from the boundary
conditions (\ref{KG_A}), (\ref{KG_B}), and functions $a_{-}(x,\theta )$ and $%
b_{+}(x,\theta )$ are obtained by integration along the trajectories going
out of the diffusive layer. The coefficients $F$, $F^{+}$ and $G$ in (\ref%
{KG_A}), (\ref{KG_B}) and the spatial dependence of the order parameter $%
\Delta (x)$ are determined in an iterative self-consistent way using Eqs. (%
\ref{KG_A})-(\ref{F_av3}) and (\ref{Sc0a})-(\ref{Sc0c}), respectively. All
the calculations below were performed at temperature $T=0.5T_{C}$.

According to this procedure, we calculate spatial distributions of pair
potential $\Delta (x)$, pair amplitude $f(x)$ and surface DoS (SDoS) for
different thicknesses of diffusive layer.

Figure \ref{Del_x} shows spatial dependencies of the pair potential $\Delta
(x)$ for $p_{x}$ and chiral $p_{x}+ip_{y}$ cases. In a $p_{x}$-wave
superconductor, the amplitude of the pair potential $\Delta _{x}$ reaches
its maximum value in the bulk ($\Delta _{x}\approx 2T_{C}$ at $T=0.5T_{C}$).
In the vicinity of the interface it is suppressed up to zero in the absence
of a diffusive layer. It comes from the fact that the reflection of
electrons takes place into the band with negative sign of pair potential
(See Fig. \ref{Del_x}a). The presence of roughness does not change the
general shape of the dependence and only provides slight growth of the pair
potential $\Delta _{x}(0)$ at the surface (Figs. \ref{Del_x}c and \ref%
{Del_dl}a).

\begin{figure}[tbh]
\begin{center}
\includegraphics[width=8.5cm]{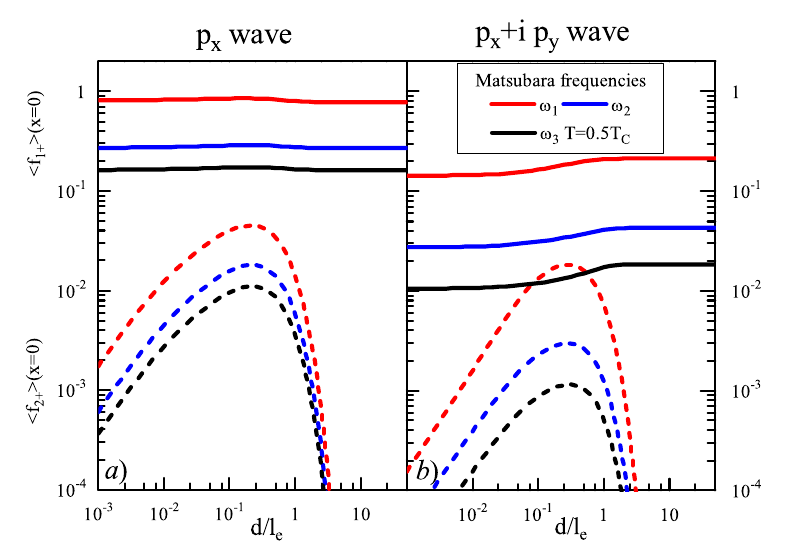}
\end{center}
\caption{Odd $<f_{1}>$ and even $<f_{2}>$ components of pair amplitude at a
surface ($x=0$) versus its roughness $d/l_{e}$. Panels a) and b) correspond
to $p_{x}$-wave and to $p_{x}+ip_{y}$-wave respectively. Pair amplitudes are
angle averaged and calculated for the first, the second and third Matsubara
frequencies $\protect\omega _{n}$ at temperature $T=0.5T_{C}$. }
\label{even_dl}
\end{figure}

\begin{figure}[tbh]
\begin{center}
\includegraphics[width=8.5cm]{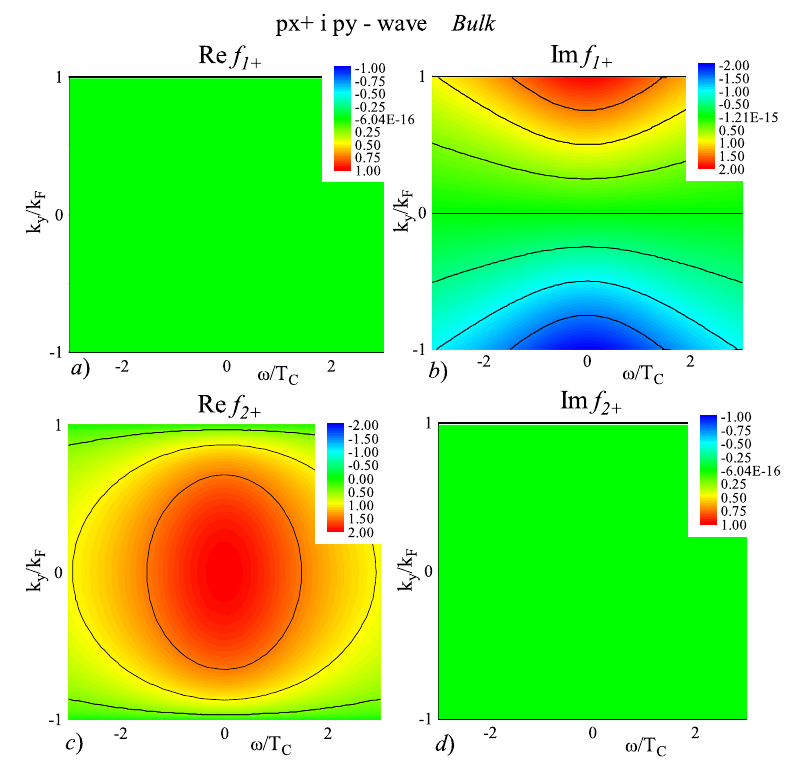}
\end{center}
\caption{(Color Online) Odd and even angle resolved pairing amplitudes $f_{1}
$ and $f_{2}$ in the bulk chiral superconductor as functions of $k_{y}$ and $%
\protect\omega $.}
\label{f_pxpy_bulk}
\end{figure}

For the case of chiral symmetry, the impact of surface is more diverse (See
Fig. \ref{Del_x}b,d). In contrast to the former case, the bulk pair
potential has the BCS magnitude (for the considered temperature $\Delta
_{x}=\Delta _{y}\approx 1.67T_{C}$) due to spherical symmetry of $\left\vert
\Delta \right\vert $. As in the previous case, the component $\Delta _{x}$
is suppressed in the vicinity of a surface. In contrary, the component $%
\Delta _{y}$ grows up to the bulk value for $p_{y}$ symmetry in the case of
a clean surface. However, $\Delta _{y}$ is sensitive to a degree of surface
roughness: the pair potential component $\Delta _{y}$ decreases by about
three times in comparison with bulk value in the limit of large roughness
(Fig. \ref{Del_dl}b). This property has a simple qualitative explanation: in
the clean limit the incident and reflected electrons fill the same sign of
pair potential, while in the diffusive case some of the reflected electrons
fills the opposite sign of the pair potential due to impurity scattering. In
the following we will see that this phenomenon manifests itself in the DoS
at a surface.

\begin{figure}[tbh]
\begin{center}
\includegraphics[width=8.5cm]{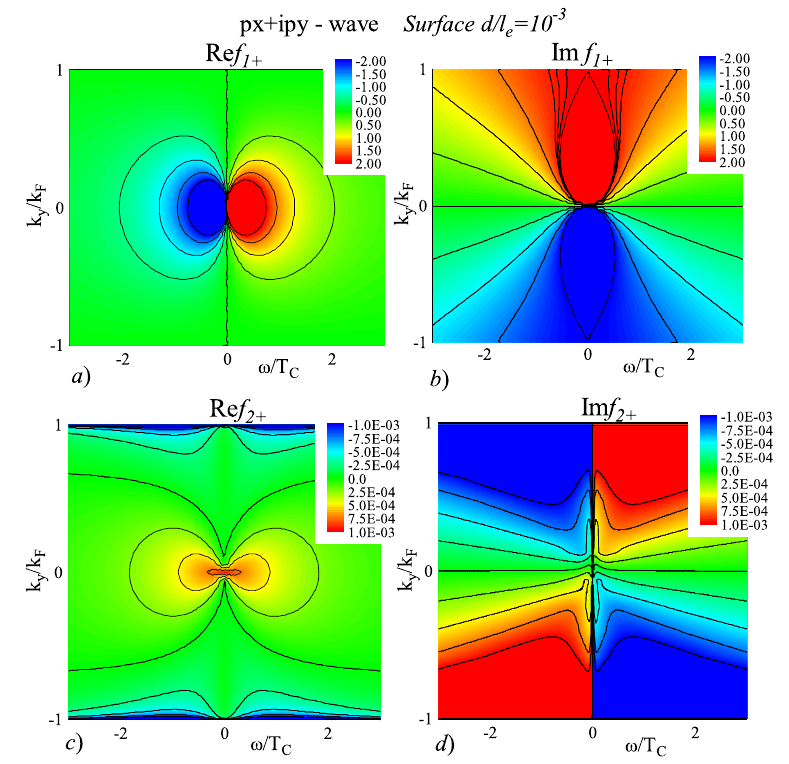}
\end{center}
\caption{(Color Online) Odd and even angle resolved pairing amplitudes $f_{1}
$ and $f_{2}$ as functions of $k_{y}$ and $\protect\omega $ at the almost
mirror surface ($x=0$) with roughness ($d/l_{e}=0.001$) for the chiral
superconductor.}
\label{f_pxpy_1e3}
\end{figure}

\begin{figure}[tbh]
\begin{center}
\includegraphics[width=8.5cm]{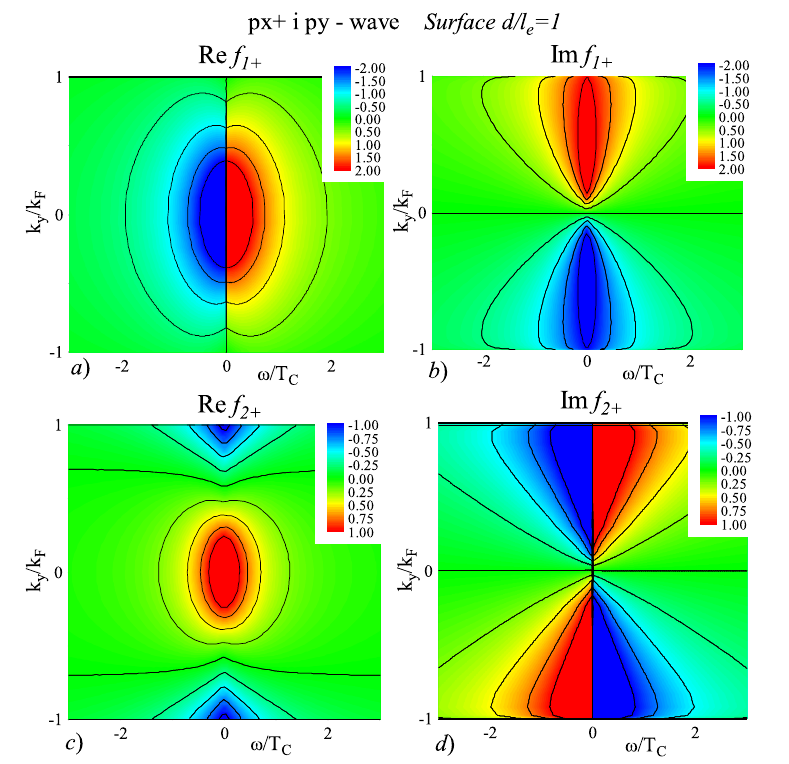}
\end{center}
\caption{(Color Online) Odd and even angle resolved pairing amplitudes $f_{1}
$ and $f_{2}$ as functions of $k_{y}$ and $\protect\omega $ at the surface ($%
x=0$) with roughness ($d/l_{e}=1$) for the chiral superconductor.}
\label{f_pxpy_1}
\end{figure}

\section{\protect\bigskip Pair amplitudes $f$ and $f^{+}$\label{Sec3}}

An important characteristic of the considered system is a relation between
surface roughness and the time-parity of the pairing amplitude $f$ near the
surface. Let us introduce the symmetrized functions%
\begin{equation}
f_{1\pm }=f_{\pm }-f_{\pm }^{+},
\end{equation}%
\begin{equation}
f_{2\pm }=f_{\pm }+f_{\pm }^{+}.
\end{equation}%
As follows from Eqs. (\ref{El0})-(\ref{El0b}) (see Appendix B), these Green
functions have the following symmetries with respect to the Matsubara
frequency:

\begin{equation}
g_{\omega }=-g_{-\omega }^{\ast };\ \ f_{1,\omega }=-f_{1,-\omega }^{\ast
};~~f_{2,\omega }=f_{2,-\omega }^{\ast }  \label{RelF}
\end{equation}%
and with respect to the angle of motion $\theta $

\begin{equation}
f_{1}(\theta )=f_{1}^{\ast }(-\theta );~~f_{2}(\theta )=f_{2}^{\ast
}(-\theta );~~g(\theta )=g^{\ast }(-\theta ).  \label{RelA}
\end{equation}%
Such symmetry also means that imaginary parts of these functions are
antisymmetric over $\theta $ and disappear after averaging over $\theta .$
Therefore, the average quantities 
$<f_{1,2}>$ are real functions and we can call function 
$<f_{1}>$ odd-frequency and 
$<f_{2}>$ even-frequency. To demonstrate this property, we trace the
behavior of 
$<f_{1}>$ and 
$<f_{2}>$ in detail in the bulk superconductor and at the surface.

For a $p_{x}$-wave superconductor, the problem can be simplified and can be
solved in terms of real values: the function $f_{1}(f_{2})$ is symmetric
over angle $\theta $ and odd (even) over frequency $\omega $. We will focus
on the functions $f_{1+}$ and $f_{2+}$ corresponding to incident
trajectories. In the bulk superconductor only the even-frequency component $%
f_{2+}$ exists in full accordance with analytical solutions (\ref{As0})-(\ref%
{As1}) (Fig. \ref{f_px}a,b). Hereinafter, we will present angle dependencies
in terms of parallel component of the Fermi wave-vector $k_{y}=k_{F}\sin
(\theta )$. At the surface the formation of another component takes place:
electrons reflect into the lobe with different sign of order parameter (in
accordance with Eqs. (\ref{BC1a})-(\ref{BC1b})) and an odd-frequency Green
function $f_{1+}$ (Fig. \ref{f_px}c) is generated. Its amplitude diverges in
the limit $\omega \rightarrow 0$, but remains finite at a certain Matsubara
frequency $\omega _{n}$.

The behavior of even-frequency $f_{2+}$ is a quite complex. At the mirror
surface it is fully destroyed by direct reflection of particles in
accordance with Eqs. (\ref{BC1a})-(\ref{BC1b}). Surface roughness leads to
generation of even-frequency Green function $f_{2+}$ since reflected
amplitudes $a_{-}$ and $b_{+}$ become isotropic. However the average value $%
\langle f_{1+}\rangle $ during further isotropization reaches its maximum
and starts to decrease for larger roughness values (See Fig. \ref{even_dl}%
a). This effect occurs because $f_{2+}$ has different signs at angles $%
\theta $ in the vicinity of $\pm \pi /2$ and $\theta =0$ and in the limit of
a thick diffusive layer these angle areas compensate each other during
integration.

\begin{figure}[tbh]
\begin{center}
\includegraphics[width=8.5cm]{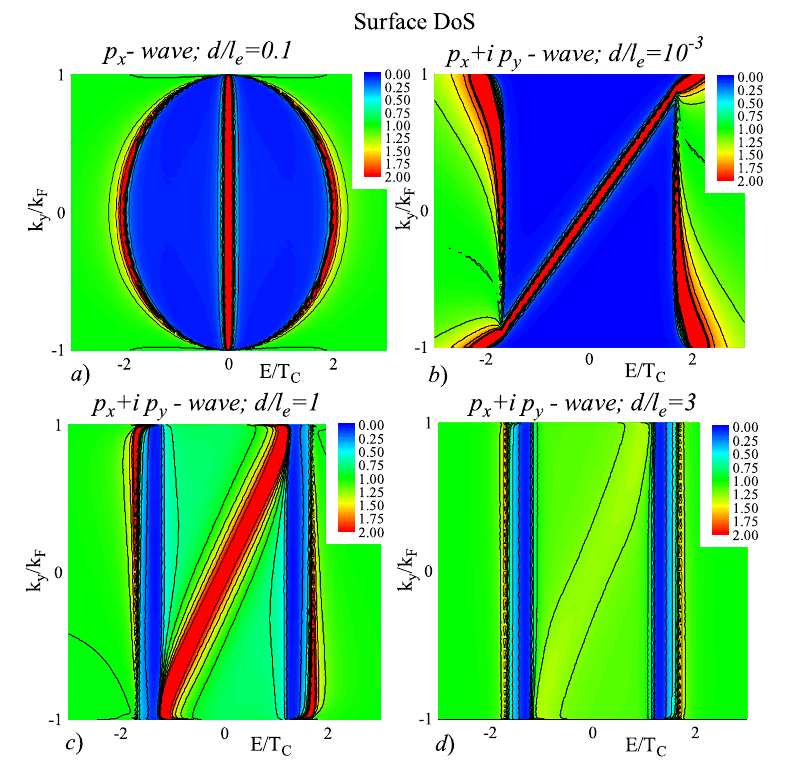}
\end{center}
\caption{(Color Online) Surface $x=d$ angle-resolved DoS for a) $p_{x}$-wave
with surface roughness $d/l_{e}=0.1$ and for b), c), d)$p_{x}+ip_{y}$-wave
with roughness $d/l_{e}=10^{-3}$, $d/l_{e}=1$, $d/l_{e}=3$ respectively.}
\label{SARDOS3d}
\end{figure}


In the chiral $p_{x}+ip_{y}$ -wave superconductors the general properties of
the Green functions are pretty similar: in the bulk all odd-frequency
components of Green functions Re $f_{1}$ and Im $f_{2}$ don't exist (Fig. %
\ref{f_pxpy_bulk}a,d) and even ones correspond to the symmetry of real and
imaginary parts of pair potential $\Delta $. Thus Re $f_{2+}$ has maximum at 
$\theta =0$, in accordance with the angle-dependence of $p_{x}$-component,
and Im$f_{1+}$ reaches its maximum values at $\theta =\pi /2$ in accordance
with $ip_{y}$-one. (Fig. \ref{f_pxpy_bulk}b,c)

In the vicinity of the surface other components also arise. Particles
reflected from the mirror boundary into the $p_{x}$-band with different sign
of order parameter generate an odd-frequency pair amplitude. However, in
imaginary values the sign of the $p_{y}$-component of the order parameter is
conserved after reflection and hold even-frequency symmetry. Thus in this
case there are only two significant components of Green functions:
odd-frequency Re $f_{1+}$ with maximum at ($\theta =0$) and even-frequency
Im $f_{1+}$ increasing for large angles. (See Fig. \ref{f_pxpy_1e3}). In the
structures with finite thickness of diffusive layer another Green function
components arise. Isotropization of $a_{-}$ and $b_{+}$ leads to the
formation of nonzero components Re $f_{2}$ and Im $f_{2}$. At greater
roughness they increase further (Fig. \ref{f_pxpy_1}), but the averaged
value of $\left\langle f_{2+}\right\rangle $ falls down due to negative
contribution from large angles.

To show this clearly we present angle averaged pair amplitudes $%
|\left\langle f_{1+}\right\rangle |$ and $|\left\langle f_{2+}\right\rangle
| $ at the surface versus roughness (Fig. \ref{even_dl}) for the first, the
second and the third Matsubara frequencies at fixed temperature $T=0.5T_{C}$%
. Odd-frequency amplitude $|\left\langle f_{1+}\right\rangle |$
significantly exceeds even-frequency one $|\left\langle f_{2+}\right\rangle
| $ in cases of both $p_{x}$ and chiral $p_{x}+ip_{y}$ symmetries.
Furthermore, in limits of both low and high roughness the even-frequency
component $|\left\langle f_{2+}\right\rangle |$ vanishes. At the same time ,
we have found that this component reaches its maximum value in the finite
roughness range. This means that new effects exist in the range of
intermediate roughness and one may expect a qualitative difference in
measurable properties such as DoS in this regime.

\section{Density of States\label{Sec4}}

To calculate DoS, one can solve the same system of equations (\ref%
{El_ric_cl_A})-(\ref{BC1b}), where Matsubara frequency is replaced by energy 
$\omega \rightarrow iE$. Further we will focus only on DoS for incident
electrons because it is this quantity which is probed in tunnel experiments

\begin{equation}
DoS=Re(g_{+})=Re\left( \frac{1+a_{+}b_{+}}{1-a_{+}b_{+}}\right)
\end{equation}

\begin{figure}[tbh]
\begin{center}
\includegraphics[width=8.5cm]{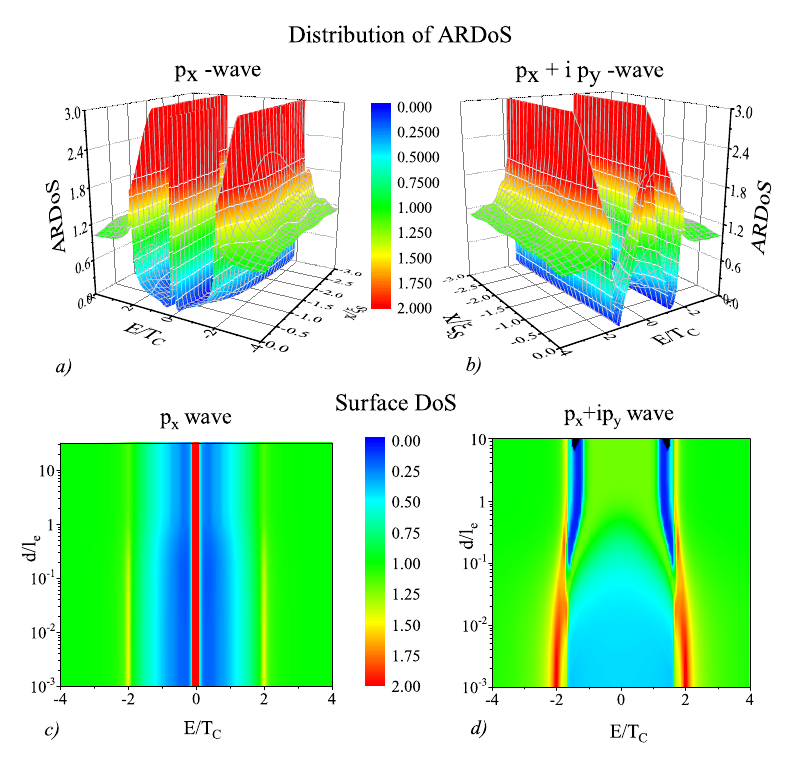}
\end{center}
\caption{(Color Online) a,b) Spatial distribution of ARDoS ($\protect\theta %
=0$) for the $p_{x}$-wave and $p_{x}+ip_{y}$-wave superconductors. c-d)
Angle Averaged SDoS as a function of surface roughness.}
\label{Last_Figure}
\end{figure}

Odd-frequency pairing around the surface leads to formation of subgap bound
states, which occur as peaks in the angle resolved density of states (ARDoS)
in both $p_{x}$ and $p_{x}+ip_{y}$ cases. Figure \ref{SARDOS3d} shows ARDoS
at the surface for $p_{x}$-wave (a) and for chiral superconductor (b, c, d)
and reveals the behavior of the subgap bound states as a function of angle
of propagation $\sin (\theta )$ and momentum $k_{y}$, respectively. In the $%
p_{x}$-case the peak is narrow and keeps its zero energy position for every $%
k_{y}$. The width of the gap is determined by the bulk pair potential $%
\Delta (\theta )=\Delta _{x}\cos (\theta )$ despite the pair potential at
the surface is almost absent. Therefore, the predominant contribution to
formation of ARDoS at the surface is provided by the proximity effect with
the bulk superconductor.

In contrast, for the chiral symmetry case (Fig.\ref{SARDOS3d}b-d), the
energy of a bound state depends linearly on $k_{y}$. The dispersion of the
corresponding peak depends on surface properties: the higher the roughness,
the wider this peak. The value of the gap in the surface DoS is now $k_{y}$
independent and is also determined by proximity with the bulk material.
However, for high $k_{y}$ (for the particles moving almost parallel to the
surface) it grows up to $\Delta _{y}$ at the surface. In accordance with
Fig. \ref{Del_x} it provides different properties in the limits of clean and
rough surface since the value of $\Delta _{y}$ can be larger or smaller
compared to the bulk. All these effects appear in the vicinity of the
surface at distances of the order of coherence length (Fig. \ref{Last_Figure}%
a-b).

\begin{figure}[tbh]
\begin{center}
\includegraphics[width=8.5cm]{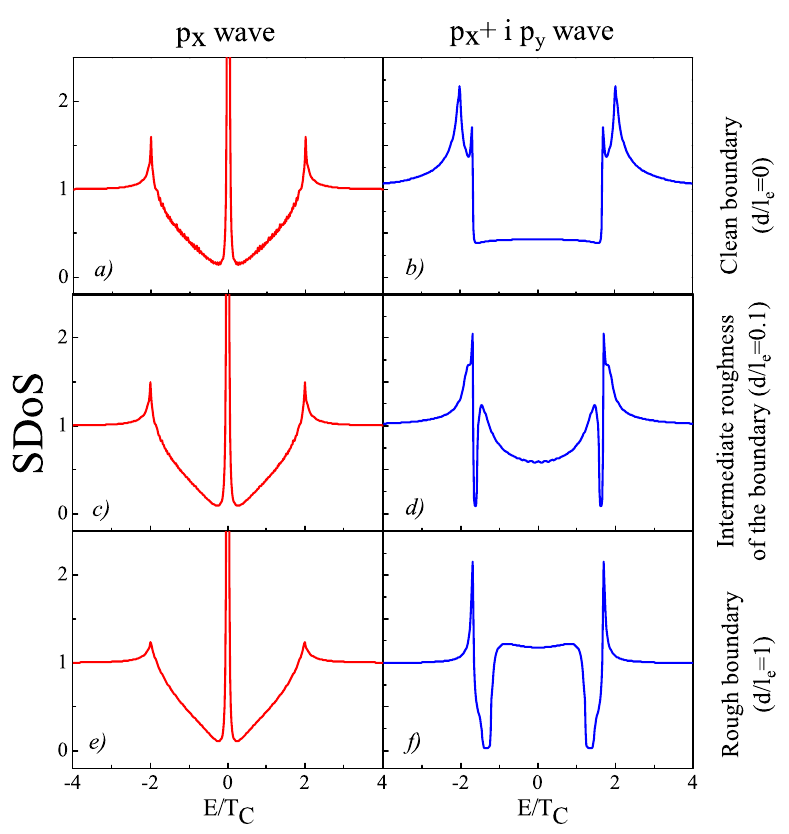}
\end{center}
\caption{Angle Averaged Surface DoS for various values of the roughness
parameter: ($d/l_{e}=0$ at the upper panels, $d/l_{e}=0.1$ at the middle
panels and $d/l_{e}=1$ at the lower panels). Left and right columns of
panels correspond to the $p_{x}$ and $p_{x}+ip_{y}$ symmetry, respectively}
\label{SDoS}
\end{figure}

These properties lead to totally different angle averaged DoS for various
symmetries. $p_{x}$ superconductor preserves the zero energy peak and peaks
at the bulk $\Delta $ even in an averaged surface DoS (SDoS)\ regardless of
the roughness (Fig. \ref{SDoS}a,c,e). The results for the chiral case
strongly depend on properties of the surface. In the case of mirror surface
(Fig. \ref{SDoS}b) angle-averaged subgap SDoS transforms into plateau with a
value around zero. The structure above the gap includes two peaks: the
intrinsic ($E=2T_{C}$) one provided by the surface component $\Delta _{y}$
of the pair potential and the proximity one ($E=1.67T_{C}$), which appears
due to influence of the bulk part of material. However, the growth of
roughness leads to suppression of $\Delta _{y}$ and to the shift of
intrinsic peak inside the energy range between proximity peaks (Fig. \ref%
{SDoS}d). At the same time, the magnitude of the middle plateau grows until
it merges with intrinsic peaks in the limit of dirty surface (Fig. \ref{SDoS}%
f). It also provides formation of DoS dips between proximity and intrinsic
peaks. Dependence of the surface DoS on roughness $d/l_{e}$ is presented in
Fig. \ref{Last_Figure}c-d, where it is demonstrated how the intrinsic peak
shifts into the gap with an increase of roughness.

\section{Conclusion}

In this work we have derived effective boundary conditions at diffusive
surface of clean $p$-wave superconductor. Using the developed approach, we
study both $p_{x}$ and $p_{x}+ip_{y}$ - wave superconductors with various
surface properties ranging from the mirror to heavily rough. We consider the
behavior of the most important characteristics of these systems: pair
potential $\Delta $, pair amplitudes $f_{1}$ and $f_{2}$ and density of
states as a function of surface roughness. In the $p_{x}$ case we
demonstrate the robustness of zero-energy peak in the density of states with
respect to surface roughness. This effect is due to stability of
odd-frequency pairing state at the surface with respect to disorder. In the
case of chiral $p_{x}+i~p_{y}$ state we demonstrate the appearance of
complex multi-peak subgap structure with increasing surface roughness.
Furthermore, the systems with a finite surface roughness provide more
complicated spectra than in the limits of mirror or heavily rough surfaces.
This fact should be taken into account in interpretation of the results of
tunneling spectroscopy of unconventional superconductors.

Finally, it is important to note that the robust zero-energy peak in DoS
discussed in this work is protected by topology. For example, the
topological origin of the flat band on the surface of a d-wave
superconductor has been clarified in \cite{Ryu} (see further referencers in 
\cite{Silaev}). Topologial stability of surface bound states of
two-dimensional $p_x$ wave and chiral p-wave superconductors has been
studied in \cite{Furusaki,Sato}.

\begin{acknowledgments}
The work is supported by the Russian Federation Basic Research Foundation,
grant no. 13-02-01085 (M.Yu.K), by the Ministry of Education and Science of
the Russian Federation, grant no. 14Y26.31.0007 and Scholarship of the
President of the Russian Federation. One of the authors (Y.T.) is supported
by Grants-in-Aid for Scientific Research from the Ministry of Education,
Culture, Sports, Science and Technology of Japan Topological Quantum
Phenomena (Grant No. 22103005 ) and the Strategic International Cooperative
Program (Joint Research Type) from the Japan Science and Technology Agency.
\end{acknowledgments}

\bigskip

\appendix

\section{Diffusive layer solution \label{Appendix}}

The solution of equations (\ref{El_ric_dr_A})-(\ref{El_ric_dr_B}) in the
diffuse layer can be represented as \cite{Golubov2}%
\begin{equation}
\frac{F^{+}a_{\pm }+G-1}{F^{+}a_{\pm }+G+1}=C_{1\pm }\exp \left\{ \mp
kx\right\} ,  \label{sola}
\end{equation}%
\begin{equation}
\frac{Fb_{\pm }+G-1}{Fb_{\pm }+G+1}=C_{2\pm }\exp \left\{ \pm kx\right\}
,\quad  \label{solb}
\end{equation}%
where $C_{1\pm }$ and $C_{2\pm }$ are integration constants 
\begin{equation}
k=\frac{\sqrt{\left\langle g_{+}+g_{-}\right\rangle ^{2}+\left\langle
f_{+}^{+}+f_{-}^{+}\right\rangle \left\langle f_{+}+f_{-}\right\rangle }}{%
\ell \cos (\theta )},  \label{Eq80}
\end{equation}%
\begin{equation*}
F=\frac{\left\langle f_{+}+f_{-}\right\rangle }{\sqrt{\left\langle
g_{+}+g_{-}\right\rangle ^{2}+\left\langle f_{+}^{+}+f_{-}^{+}\right\rangle
\left\langle f_{+}+f_{-}\right\rangle }},
\end{equation*}%
\begin{equation*}
F^{+}=\frac{\left\langle f_{+}^{+}+f_{-}^{+}\right\rangle }{\sqrt{%
\left\langle g_{+}+g_{-}\right\rangle ^{2}+\left\langle
f_{+}^{+}+f_{-}^{+}\right\rangle \left\langle f_{+}+f_{-}\right\rangle }}
\end{equation*}%
\begin{equation}
G=\frac{\left\langle g_{+}+g_{-}\right\rangle }{\sqrt{\left\langle
g_{+}+g_{-}\right\rangle ^{2}+\left\langle f_{+}^{+}+f_{-}^{+}\right\rangle
\left\langle f_{+}+f_{-}\right\rangle }}.  \label{Eq81}
\end{equation}%
Integration constants $C_{1\pm }$ and $C_{2\pm }$ can easily be expressed in
terms of functions $b_{\pm }$ and $a_{\pm }$ on the boundary of the
diffusion layer with a superconductor $(x=0)$. So for the constant $C_{1\pm
} $ and $C_{2\pm }$ it is possible to get 
\begin{equation}
C_{2-}=\frac{Fb_{-}(0,\theta )+G-1}{Fb_{-}(0,\theta )+G+1},  \label{Eq84}
\end{equation}%
\begin{equation}
C_{1+}=\frac{F^{+}a_{+}(0,\theta )+G-1}{F^{+}a_{+}(0,\theta )+G+1}.
\label{Eq85}
\end{equation}%
Substituting (\ref{Eq84}), (\ref{Eq85}) into the solution (\ref{sola}), (\ref%
{solb}) for the functions $b_{-}(d,\theta )$ and $a_{+}(d,\theta )$ at the
free surface of the diffusion layer we get

\begin{equation}
b_{-}(d,\theta )=-\frac{G}{F}+\frac{1}{F}\frac{1+p}{1-p},\ p=C_{2-}\exp
\left\{ -kd\right\} ,  \label{Eq90}
\end{equation}%
\begin{equation}
a_{+}(d,\theta )=-\frac{G}{F^{+}}+\frac{1}{F^{+}}\frac{1+q}{1-q},\
q=C_{1+}\exp \left\{ -kd\right\} .  \label{Eq91}
\end{equation}%
Proceeding in a similar way, it is easy to see that 
\begin{equation}
a_{-}(0,-\theta )=-\frac{G}{F^{+}}+\frac{1}{F^{+}}\frac{1+u}{1-u},
\label{ap}
\end{equation}%
\begin{equation}
b_{+}(0,-\theta )=-\frac{G}{F}+\frac{1}{F}\frac{1+v}{1-v},  \label{bm}
\end{equation}%
where 
\begin{equation}
u=\exp \left\{ -kd\right\} \frac{F^{+}a_{+}(d,\theta )+G-1}{%
F^{+}a_{+}(d,\theta )+G+1}=q\exp \left\{ -kd\right\} ,  \label{u}
\end{equation}%
\begin{equation}
v=\exp \left\{ -kd\right\} \frac{Fb_{-}(d,\theta )+G-1}{Fb_{-}(d,\theta )+G+1%
}=p\exp \left\{ -kd\right\} .  \label{v}
\end{equation}%
The resulting equations (\ref{Eq84})-(\ref{v}) and boundary conditions (\ref%
{BC1a}), (\ref{BC1b}) set the desired relation between the functions of a
coming $a_{+}(0,-\theta )$, $b_{-}(0,-\theta )$ and leaving $a_{-}(0,\theta
) $, $b_{+}(0,\theta )$ of the diffusion layer 
\begin{equation}
a_{-}(0,\theta )=\frac{a_{+}(0,-\theta )-\left( Ga_{+}(0,-\theta )-F\right)
\tanh \left\{ kd\right\} }{\left( F^{+}a_{+}(0,-\theta )+G\right) \tanh
\left\{ kd\right\} +1},  \label{ap0}
\end{equation}%
\begin{equation}
b_{+}(0,\theta )=\frac{b_{-}(0,-\theta )-\left( Gb_{-}(0,-\theta
)-F^{+}\right) \tanh \left\{ kd\right\} }{\left( Fb_{-}(0,-\theta )+G\right)
\tanh \left\{ kd\right\} +1}.  \label{bm0}
\end{equation}%
From relations (\ref{ap0}), (\ref{bm0}), it follows that (as in d-wave case 
\cite{Golubov2}) the values of the modified functions Eilenberger on leaving
the border trajectory can be divided into two parts. One of them 
\begin{equation}
a_{-,d}(0,\theta )=\frac{F}{1+G},\ b_{+,d}(0,\theta )=\frac{F^{+}}{1+G}
\label{apd}
\end{equation}%
is determined by the uncorrelated contribution to the direction of the angle 
$\theta $. It is formed as a result of rescattering in this corner of
particles incident on the diffuse layer in the whole range of trajectories
towards this layer. It is easy to see that this part defines the functions
of $a_{-}(0,\theta )$ and $b_{+}(0,\theta )$ in the limit of large thickness
of the diffusion layer, $kd\gg 1.$ In this case, the electrons incident and
reflected from the surface are completely uncorrelated. The remaining parts 
\begin{equation}
b_{+,c}(0,\theta )=\frac{\left( 1-\tanh \left( kd\right) \right) \left(
\left( 1+G\right) b_{-}(0,-\theta )-F^{+}\right) }{\left( 1+G\right) \left(
\left( Fb_{-}(0,-\theta )+G\right) \tanh \left\{ kd\right\} +1\right) },
\label{bmc}
\end{equation}%
\begin{equation}
a_{-,c}(0,\theta )=\frac{\left( 1-\tanh \left( kd\right) \right) \left(
\left( 1+G\right) a_{+}(0,-\theta )-F\right) }{\left( 1+G\right) \left(
\left( F^{+}a_{+}(0,-\theta )+G\right) \tanh \left\{ kd\right\} +1\right) },
\label{apc}
\end{equation}%
set the degree of correlation between the incoming and outgoing from the
boundary trajectories. It is evident that this correlation is stronger, the
smaller the thickness of the diffusion layer $d$. Indeed, from (\ref{bmc}), (%
\ref{apc}), it follows that at angles 
\begin{equation*}
\frac{\pi }{2}\leq \theta \leq \arccos (\frac{d\sqrt{\left\langle
g_{+}+g_{-}\right\rangle ^{2}+\left\langle f_{+}^{+}+f_{-}^{+}\right\rangle
\left\langle f_{+}+f_{-}\right\rangle }}{l})
\end{equation*}%
scattering is mainly diffusive. With decreasing thickness, this region of
angles shrinks, so that in the limit of small thickness $(kd\ll 1)$ it is
more and more limited by trajectories, moving along the border. As a rule,
they do not contribute to physical observables (DoS, the conductance, the
critical current of Josephson junctions). For all other paths that define
these values, the boundary conditions (\ref{ap0}), (\ref{bm0}) reduce in
this limit to the mirror (\ref{BC1a}), (\ref{BC1b}) type.

\section{Symmetry relations \label{Symmetry}}

The system of Eilenberger equations is

\begin{gather}
2\omega f_{\omega }(x,\theta )+v\cos (\theta )\frac{d}{dx}f_{\omega
}(x,\theta )=2\Delta g_{\omega }(x,\theta ), \\
2\omega f_{\omega }^{+}(x,\theta )-v\cos (\theta )\frac{d}{dx}f_{\omega
}^{+}(x,\theta )=2\Delta ^{\ast }g_{\omega }(x,\theta ), \\
2v\cos (\theta )\frac{d}{dx}g_{\omega }(x,\theta )=2\left( \Delta ^{\ast
}f_{\omega }-\Delta f_{\omega }^{+}\right) . \\
\Delta =\Delta _{x}\cos \theta +i\Delta _{y}\sin \theta
\end{gather}
First we consider relations with respect to $\theta .$ We write them for
angle $-\theta $ and conjugate, resulting in

\begin{gather}
2\omega f_{\omega }^{\ast }(-\theta )+v\cos (\theta )\frac{d}{dx}f_{\omega
}^{\ast }(-\theta )=2\Delta g_{\omega }^{\ast }(-\theta ), \\
2\omega f_{\omega }^{+\ast }(-\theta )-v\cos (\theta )\frac{d}{dx}f_{\omega
}^{+\ast }(-\theta )=2\Delta ^{\ast }g_{\omega }^{\ast }(-\theta ), \\
2v\cos (\theta )\frac{d}{dx}g_{\omega }^{\ast }(-\theta )=2~(\Delta ^{\ast
}f_{\omega }^{\ast }(-\theta )-\Delta f_{\omega }^{+\ast }(-\theta )).
\end{gather}

This set of equations coinsides with initial one after the following
substitution 
\begin{gather}
g_{\omega }(\theta )=g_{\omega }^{\ast }(-\theta )  \label{th1} \\
f_{\omega }(\theta )=f_{\omega }^{\ast }(-\theta )  \label{th2} \\
f_{\omega }^{+}(\theta )=f_{\omega }^{+\ast }(-\theta )  \label{th3}
\end{gather}%
This proves angle-symmetry relations (\ref{RelA}).

\bigskip

Next, we consider symmetry of Eilenberger equations with respect to
Matsubara frequency $\omega $. Similarly to the previous step, we take
equations at negative frequncy $-\omega $ and conjugate them. After
conjugation and some rearrangements we arrive 
\begin{gather}
2\omega f_{-\omega }^{\ast }(\theta )-v\cos (\theta )\frac{d}{dx}f_{-\omega
}^{\ast }(\theta )=2\Delta ^{\ast }\left( -g_{-\omega }^{\ast }(\theta
)\right) , \\
2\omega f_{-\omega }^{+\ast }(\theta )+v\cos (\theta )\frac{d}{dx}f_{-\omega
}^{+\ast }(\theta )=2\Delta \left( -g_{-\omega }^{\ast }(\theta )\right) , \\
2v\cos (\theta )\frac{d}{dx}\left( -g_{-\omega }^{\ast }(\theta )\right)
=2~(\Delta ^{\ast }f_{-\omega }^{+\ast }-\Delta f_{-\omega }^{\ast }).
\end{gather}%
Comparison with initial equations provides the required symmetry relations (%
\ref{RelF}).%
\begin{gather}
g_{-\omega }(\theta )=-g_{\omega }^{\ast }(\theta )  \label{w1} \\
f_{-\omega }(\theta )=f_{\omega }^{+\ast }(\theta )  \label{w2} \\
f_{-\omega }^{+}(\theta )=f_{\omega }^{\ast }(\theta )  \label{w3}
\end{gather}

\end{document}